# Exchange-driven Collective Behavior in a 3D Array of Nanoparticles

Ha M. Nguyen* and Pai-Yi Hsiao†

*Department of Engineering and System Science,*

*National Tsing Hua University, Hsinchu, Taiwan 30013, R.O.C*

(Dated: November 11, 2018)

0

## Abstract

A Monte Carlo simulation is performed in a cubic lattice of interacting identical Stoner-Wohlfarth nanoparticles. The model system is a randomly-anisotropic Heisenberg spin system with a small anisotropy-to-exchange ratio $D/J = 3.5$. The dc susceptibility, $\chi_{\text{dc}}(T)$, shows a Curie-Weiss-like transition at a temperature $T_{\text{C}}/J \approx 1.5$, followed by a low-temperature glassy behavior manifested by cusps in both the zero-field-cooled and the field-cooled curves. The ac susceptibility, $\chi_{\text{ac}}(T,\omega)$, at various frequencies, $\omega$, shows that with decreasing temperature, a non-Arrhenius dispersive peak occurs at $T_{\text{b}}(\omega)$, succeeded by another dispersionless peak at $T_{\text{g}}/J \approx 1.20$ in the in-phase part, $\chi'(T,\omega)$, of $\chi_{\text{ac}}(T,\omega)$ while the out-of-phase part, $\chi''(T,\omega)$, shows only one peak. A dynamic scaling analysis shows that the system exhibits a critical slowing-down at $T_{\text{g}}$ with a quite small exponent $z\nu \approx 1.65$. However, no universal collapse is seen for the fully-scaled data of $\chi''(T,\omega)$. These observed behaviors are interpreted under the droplet-like hypothesis that the formation and development of exchange-induced correlated clusters drive ensembles of nanoparticles undergoing a transition from a paramagnetic order to a short-range order (SRO) at $T_{\text{C}}$, followed by a transition at $T_{\text{g}}$ to the magnetic state in which a magnetic glassy order and a magnetic quasi-long-range order (QLRO) coexist. In addition, our simulation shows that the onset of the latter transition, which is peculiarly manifested by the dispersionless peak, occurs only for those ensembles possessing the anisotropy strength in the region $1.0 \leq D/J \leq 5.0$. When this range is exceeded, this onset totally suppressed. The reason that the QLRO is prohibited in ensembles of strong random anisotropy may account for this phenomenon.

*Electronic address: nmha@ess.nthu.edu.tw;

Permanent address: Institute of Materials Science, Vietnamese Academy of Science and Technology, Hanoi, Vietnam

†Electronic address: pyhsiao@ess.nthu.edu.tw

In 1973, Harris, Plischke, and Zuckermann (HPZ) [1] proposed the so-called random anisotropy model (RAM) for magnetism in rare-earth amorphous compounds because of the fact that both the Curie temperature $T_\text{C}$ and the spontaneous magnetization $M_\text{s}$ of the TbFe$_2$ amorphous alloy are substantially lower than those of its crystalline counterpart observed by Rhyne, Pickart, and Alperin [2]. Since then, the model has been considered as the prototype for the physics of magnetic phenomena, not only in magnetic amorphous alloys but also in other disordered magnetic materials possessing random anisotropy [3, 4, 5, 6], among which the RAM has been subjected to various developments in modeling ensembles of magnetic nanoparticles (NPs) over the past decades [7, 8, 9, 10].

In this paper, we will apply the RAM to explore, using a Monte Carlo (MC) simulation, the nature of low-temperature magnetic collective behavior driven by the competition between the exchange interaction and the random anisotropy in an NP array. The studied system is a simple cubic lattice of NPs, which is the simplest model among 3D ordered nanostructures [11, 12, 13]. Our simulation is carried out under the RAM in the sense that the magnetic moments of same-sized Stoner-Wohlfarth (SW) NPs [14] play a role as atomic spins in the original RAM model [1]. The magnetic moment can be represented as a vector $\vec{\mu} = M_\text{s} V \hat{\mu}$, where $M_\text{s}$ is the spontaneous moment and is assumed to be temperature-independent, $V$ is the particle volume, and $\hat{\mu}$ is the unit vector along the moment direction. The random anisotropy is that of a single particle with a positive strength $D = KV$, where $K$ is the anisotropy constant [11, 12, 13]. Therefore, the RAM's Hamiltonian in our study can be expressed as

$$\mathcal{H} = -D \sum_i (\hat{n}_i \cdot \hat{\mu}_i)^2 - J \sum_{\langle i,j \rangle} \hat{\mu}_i \cdot \hat{\mu}_j - H \sum_i \hat{\mu}_i \cdot \hat{z}, \qquad (1)$$

which is identical to that for the original model. Here, $\hat{n}$ is the unit vector along the direction of the easy-axis, $\hat{z}$ is the directional vector for the O$z$ axis of the Cartesian system (O$zyz$) along which the magnetic field with strength $H$ is applied, and $J$ is the strength of the nearest-neighbor exchange interaction, which is assumed to be positive (ferromagnetic). Notice that the terms in Eq. (1) are expressed in the convention that $D$, $J$, and $H$ are "measured" in the dimension of energy while the scalar products inside the summation signs yield dimensionless values. The temperature of the system, $T$, is in the dimension of energy, too.

The motivation for our work is associated with the recent MC work by Billoni, Cannas,

and Tamarit (BCT's work) [15] on the out-of-equilibrium dynamics of the model in Eq. (1). Interestingly, they found such universality classes for nonequilibrium spin-glass (SG) dynamics [16] as the subaging scaling law and the fluctuation-dissipation relation for a cubic lattice of Heisenberg spins. Their work is particularly significant for RAM magnetism, which is still a subject of heavy controversy related to the nature of the ordered phases at low temperatures and its dependence on the degree of anisotropy (in other words, its dependence on $D/J$) [17, 18]. The dependence on $D/J$ may be summarized briefly as follows: in strong $D/J$ region, the RAM seems to present a low-temperature spin-glass-like phase (or a speromagnetic (SPEM) phase in Coey's terminology [4]). In infinitesimally small anisotropy region, the order tends to be of locally noncollinear ferromagnetic order (also asperomagnet (ASPEM)) [4, 19]. Weak anisotropy seems to drive the ASPEM order into the so-called correlated SPEM or correlated spin glass (CSG) [19]. Recently, Itakura [20] performed a MC simulation and depicted a phase diagram for the anisotropy dependence in terms of the spin-spin correlation $G(r)$. According to that work, at low temperature, a quasi-long-range order should persist in coexistence with the CSG, at least up to an upper bound of $D/J \simeq 5.0$ with a frozen power-law $G(r) \propto r^{-\eta-1}$, above which the spins attain a short-range order (SRO) with $G(r) \propto r^{-\eta-1} \exp(-r/\xi)$. Notice that BCT's work carried out for a value of weak anisotropy strength $D/J = 3.5$ provides a strong confirmation for the Itakura's prediction of the coexistence. However, when exploring the nature of the low-temperature behavior due to the phases coexistence, questions may arise as to (i) whether the memory effects can be "observed" by these simulations and (ii) whether a SG transition temperature $T_g$ does exist. In this paper, we only address ourselves to the latter question. We find the existence of $T_g$ from a critical slowing-down scaling analysis of the ac susceptibility, $\chi_{ac}(T, \omega)$, accompanying some dynamic and ordering features characteristic of the low-temperature phases for systems of weak anisotropy. In addition, the behavior of the in-phase part of the $\chi_{ac}(T, \omega)$ at a fixed $\omega$ with various values of $D/J$ provides clues to locate the boundaries between anisotropy regions, supporting the Itakura's phase diagram.

In this work, we perform a MC simulation by using the Metropolis technique [7, 8], as in the BCT's work, for the case $D/J = 3.5$ in a cubic lattice of $L \times L \times L$ ($L = 10$) Heisenberg spins. We calculate the temperature dependence of the dc susceptibility in both the zero-field-cooled (ZFC), $\chi_{ZFC}(T)$, and the field-cooled (FC), $\chi_{FC}(T)$, protocols, and of the ac susceptibility, $\chi_{ac}(T, \omega)$. At each temperature, an initial $N_{eq} = 100\,000$-Monte-Carlo-step (MCS) run is carried out to equilibrate the system thermally, followed by another $N_{\text{av}} = 500\,000-$MCS run to get the time average in each replica. The calculated data are then configurationally averaged over a large number of replicas, $N_{\text{R}}$. The $\chi_{\text{ZFC}}(T)$ and the $\chi_{\text{FC}}(T)$ are calculated in each field protocol by using

$$\chi_{\text{ZFC,FC}}(T) = \frac{1}{H} \left[ \frac{1}{N_{\text{R}}} \sum_{R=1}^{N_{\text{R}}} \left( \frac{1}{N_{\text{av}}} \sum_{t=1}^{N_{\text{av}}} M_{\text{R}}(t) \right) \right]_{\text{T}}, \quad (2)$$

where $H$ is the strength of the dc field in the range $0.05 \leq H/J \leq 0.35$, and $M_{\text{R}}(t) = (\Sigma_{i=1}^{N} \hat{\mu}_i \cdot \hat{z})/N$ is the $\hat{z}$-component of the magnetic moment averaged over the $t$-th spin configuration of all $N = L^3$ spins after the equilibrating period (at time $0 < t \leq N_{\text{av}}$) for the $R$-th replica. To simulate the ac susceptibility, we apply a weak ac field $\vec{H}_{\text{ac}} = H_0 \sin(\omega t)\hat{z}$. The components of the ac susceptibility, $\chi_{\text{ac}}(T, \omega) = \chi'(T, \omega) - i\chi''(T, \omega)$, are calculated using the formulae in Ref. 9 as

$$\chi'(T, \omega) = \frac{1}{H_0} \left[ \frac{1}{N_{\text{R}}} \sum_{R=1}^{N_{\text{R}}} \left( \frac{1}{N_{\text{av}}} \sum_{t=1}^{N_{\text{av}}} M_{\text{R}}(t) \sin(\omega t) \right) \right]_{\text{T}}, \quad \chi''(T, \omega) = -\frac{1}{H_0} \left[ \frac{1}{N_{\text{R}}} \sum_{R=1}^{N_{\text{R}}} \left( \frac{1}{N_{\text{av}}} \sum_{t=1}^{N_{\text{av}}} M_{\text{R}}(t) \cos(\omega t) \right) \right]$$

where $H_0/J = 0.05$, and the frequency $\omega$ is varied in the range $1 \times 10^{-4} \leq \omega \leq 1 \times 10^{-1}$ (MCS$^{-1}$).

Figure 1 presents the dc magnetic "measurements" with the ZFC and the FC protocols for $H/J = 0.08$, $0.16$, and $0.32$. As one can see in Fig. 1(a), the dc susceptibility curves resemble a paramagnetic (PM) state in the high-temperature region. A least-squared fit to $1/\chi_{\text{FC}}(T)$ in this temperature region indicates a Curie-Weiss-like transition at $T_{\text{C}}/J \approx 1.50$, as shown in the inset of Fig. 1(a). This behavior causes one to think of the transition to a low-temperature ferromagnetic (FM) phase. However, the Curie-Weiss law is ruled out in the case of RAM magnetism because a FM long-range order is prohibited in the three-dimensional RAM model for any arbitrarily small value of anisotropy (i.e., $D/J \neq 0$) [17, 18]. According to Coey, one may refer the ordering phase, at sufficiently low temperatures, in the RAM to a noncollinear magnetic structure to distinguish it from the collinear magnetic structure of the FM phase [4]. Similar examples for which the Curie-Weis-law behavior cannot warrant a low-temperature FM long-range order are ensembles of strongly-interacting nanoparticles [12, 21, 22]. Also, the concept of domains becomes less obvious and unclear in RAM mangetism. For the low $D/J$ region, the domains (ASPEM domains) become extremely small for nearest-neighbor interactions. The concept of domains is lost altogether in SPEM structures (i.e., in strong $D/J$ region) because the direction of a spin changes essentially randomly from site to site, so the domain is reduced to a single site [4]. In the weak region, however, one might more or less preserve the concept of a "tiny domain" by using a possible hypothesis that at a temperature below $T_C$, the system consists entirely of small magnetic "correlated clusters" or "droplets" which are formed due to FM exchange coupling between nearest-neighbor spins. This droplet hypothesis is similar to that in spin glasses, where spin-glass domains can be renormalized to give some insight into spin-glass dynamics [12, 16]. As reported in BCT's work [15], the aging effect observed in the current case of $D/J = 3.5$ for the two-time autocorrelation function is a clear evidence strongly supporting the droplet hypothesis because the waiting-time dependence and the subaging scaling behavior are natural consequences of the droplet theory in spin-glasses [12, 16, 23]. As observed in Fig. 1(a), the clusters are probably nucleated somewhere above $T_C$, where the dc susceptibility curves for different applied fields branch ($T/J \lesssim 2.0$). This nucleating phenomenon may be described by using two competitive processes: the tendency to freeze individual spins in random orientations with decreasing temperature (hindering cluster nucleation), and the alignment of neigboring spins to form correlated clusters by exchange coupling. Therefore, $H^{-1}(\partial M_{FC}(T)/\partial T)$ may reflect the growth rate of the clusters. As shown in Fig. 1(b), the growth rate may reach its maximum, similar to $H^{-1}|\partial M_{FC}(T)/\partial T|$, at the corresponding inflection point in the $M_{FC}(T)$ curve near $T_C$, which slightly depends on $H$. The existence of SRO clusters near $T_C$ and the history-dependent nonequilibirium dynamics (e.g., aging and memory effects) both below and above the peak of $\chi_{ZFC}(T)$ have been strongly evidenced in $Y_{0.7}Ca_{0.3}MnO_3$ and $La_{0.5}Sr_{0.5}CoO_3$ compounds [24, 25]. Therefore, $T_C$ should be spoken of as a "smeared" transition temperature rather than as the Curie temperature of a PM-to-FM transition.

In the droplet theory, the response of clusters to the field can be considered analogous to the thermal activation of NPs in superparamagnets (SPMs). However, in contrast to the SPMs the distribution of cluster sizes, $P(t; V_{cluster}(t))$, is a dynamic function of their time-dependent linear sizes $L(t) \sim V_{cluster}^{1/d}(t)$, and cluster redistribution at each temperature and instant in time leads to a huge number of random energy states for droplet excitations. Therefore, some features characteristic of the droplet hypothesis at low temperatures can be distinguished from the $\chi_{ZFC}(T)$ and the $\chi_{FC}(T)$ data for spin-glass-like systems [23].

As seen in Fig. 1(a), the temperature $T_f$ at which $\chi_{\text{ZFC}}(T)$ and $\chi_{\text{FC}}(T)$ emerge with decreasing temperature almost coincides with the temperature for the $\chi_{\text{ZFC}}(T)$ peak. Another interesting observation in Fig. 1(a) is that $\chi_{\text{FC}}(T)$ displays a clear cusp below $T_f$ and decreases, together with a flat slow decrease of $\chi_{\text{ZFC}}(T)$, as the temperature is decreased. At sufficiently low temperatures, $\chi_{\text{FC}}(T)$ seems to be nearly constant. The same behavior of the dc susceptibility for conventional spin-glasses has recently been understood under the random energy model, a phase-space companion of the droplet theory, which was used to qualitatively interpret experimental results for a superspin-glass of $Fe_3N$ NPs [23].

The most interesting phenomenon for the $D/J = 3.5$ RAM that can be seen in Fig. 2(a) is the occurence of a double maximum structure in the in-phase part of the ac susceptibility, $\chi'(T,\omega)$, while the out-of-phase part, $\chi''(T,\omega)$, displays only one peak. At first glance, the shift to low temperature of the high-temperature dispersive peak, $T_b(\omega)$, with decreasing frequency seemingly resembles a SPM system of NPs, whose magnetic moments have the relaxation times that could be described by an Arrhenius behavior of Néel-Brown expression, $\tau = \tau_0 \exp(E_b/T)$. However, at temperatures not far below $T_C$, $\chi'(T,\omega)$ keeps on increasing with decreasing temperature to result in a dispersionless peak at $T_o/J \approx 1.2$. This dispersionless peak distinguishes the behavior of the ac susceptibility of the RAM from that of an ensemble of noninteracting NPs [26] and that of an ensemble of dipolar-interacting NPs [9]. As shown in the inset of Fig. 2(b), the so-called Arrhenius plot of $J/T_b(\omega)$ vs. $\log(1/\omega)$ indicates a non-Arrhenius behavior with an increasing slope. Notice that the slope in a regular Arrhenius plot (i.e., a straightline dependence) provides the value of the activation energy that spins need to surmount the energy barrier. Again, under the spirit of the droplet hypothesis, the above increase of the activation energy implies a nucleation of small clusters below $T_C$. As a result, the spins in each cluster are so correlated as to hinder free thermal activation of individual spins. The growth of clusters with decreasing temperature is probably a plausible interpretation for the increasing average energy barrier because a strong correlation among spins slows down single-spin thermal activation. Interestingly, the conventional critical slowing-down scaling

$$\tau_c = \tau^* \varepsilon^{-z\nu} = \tau^* \left(\frac{T}{T_g} - 1\right)^{-z\nu} \tag{4}$$

can work well with the frequency dependence of $T_b(\omega)$. A least-square fit of Eq. (4) to the simulated data yields the critical temperature $T_g/J \approx 1.20$, the critical exponent $z\nu \approx 1.65$, and the timescale constant $\tau^* \approx 10^2$ (MCS), as shown in Fig. 2(b). To fit Eq. (4) with the simulated data, the critical correlation time is obtained by $\tau_c = 1/\omega$, and the reduced temperature, $\varepsilon = T/T_g - 1$, is produced at each $\omega$ when $T$ takes the value of $T_b(\omega)$. It is shown that the derived value of $T_g$ by the fitting is fairly consistent with its value at the dispersionless peak of $\chi'(T,\omega)$. We believe that $T_g$ is the transition temperature to the low-temperature CSG phase in which, for weak random anisotropy, there emerges a magnetic quasi-long-range order (QLRO) as revealed in Itakura's work [20], albeit the scaled data of $\chi''(T,\omega)$ for $\varepsilon > 0$ do not overlap to the universal scaling law, $\chi''(T,\omega)/\chi_{eq}(T) = \varepsilon^\beta G(\omega \tau_c)$. The same situation has been experimentally reported for an ensemble of interacting Fe-C NPs with the concentration of 5 vol% [27], in which a satisfactory interpretation for the deviation from the scaling law was based on the cluster hypothesis. The quite small value of $z\nu \approx 1.65$ is attributed to the effect of dynamic finite-size scaling on the value $z = z_L$ in such a way that the finite-size correlation time should be written as $\tau_c(\varepsilon, L) = L^{z_L} \mathcal{F}(\varepsilon L^{1/\nu}) \propto \varepsilon^{-z_L \nu}$ rather than as in Eq. (4), which is used for real materials of infinite sizes. For this reason, Wansleben and Landau have reported that $z_L \nu \approx 1.28$ near the critical temperature of a cubic Ising ferromagnet [28], this value is quite comparable to our value.

To understand whether the dispersionless peak at $T_g$ in $\chi'(T,\omega)$ is responsible for the onset of the transition to the CSG phase in which the magnetic QLRO emerges, we study ensembles of nanoparticles with various values of $D/J$. As shown in Fig. 3(a), the $\chi'(T,\omega)$ curve at a fixed frequency, namely at $\omega = 3 \times 10^{-3}$ (MCS$^{-1}$), exhibits the dispersionless peak only for anisotropy strength in the range $1.0 \leq D/J \leq 5.0$, i.e., in the weak anisotropy region. When this region is exceeded, the peak is totally suspended. Strikingly, Itakura, as mentioned above, also found a QLRO ground state with a power-law correlation function in this weak anisotropy region [20]. QLRO was theoretically predicted to be a rare case of ordering which only exists in weak random-field and random-anisotropy systems of continuous symmetry (i.e., systems of non-Abelian symmetry) [29]. The occurence of the dispersionless peak is very likely due to the development of the QLRO at temperatures below $T_g$, which is characterized by a smooth rotation of the magnetization over the volume so that the directions of the magnetization are ferromagnetically correlated on a quite large or even infinite length, in coexistence with a magnetic glassy phase. That may be the reason why Chudnovsky *et al.* [19] have named this phase *correlated spin-glass*. Actually, we have recently carried out a formal Monte Carlo study of magnetic phase transition for a weak anisotropy strength, $D/J = 4.0$, of the RAM model. It was shown that a second order transition to the QLRO is possible [30]. It is worth noting that a similar behavior to that shown above for the $D/J = 3.5$ case has been experimentally observed in a-Ho$_{28}$Fe$_{72}$ thin film, a weak random anisotropy magnet, by Saito *et al.* [31]. The magnet exhibits not only a critical slowing down in phase transition but also a drastical broadening of the distribution of the relaxation time around $T_g$. Moreover, the authors claimed that the dispersionless peak is also observed for a sample of a-(Nd$_{1-x}$Gd$_x$)$_{28}$Fe$_{72}$ films with $x = 0$, but with increasing $x$, the peak is not observed which suggests a similar physical picture to that accounting for what is observed in Fig. 3(a).

In contrast, low-temperature magnetic behavior is different for systems of anisotropy strengths exceeding $D/J \simeq 5.0$ from that shown above for those of weak anisotropy. In Fig. 3(a), the dispersionless peak is not observed for anisotropy strengths of $D/J \geq 6.0$. This disappearance may be understood upon a result in Itakura's work [20] that the spin-spin correlation function for these systems can be well expressed as $G(r) \propto r^{-\eta-1} \exp(-r/\xi)$ with $\eta$ is an effective exponent and $\xi$ being an effective correlation length. This correlation function accounts for a short-range order which is locally attained among neighboring spins as a result of scattering spin orientation, and the absence of the dispersionless peak manifests a global distruction of the QRLO at low temperatures due to strong random anisotropy [19]. In addition, one can see clearly in Fig. 3(a) that when increasing $D/J$ the maximum of $\chi'(T,\omega)$ broadens and shifts toward the high-temperature region. The shift, however, is reversed backward to the low-temperature region when further increasing $D/J$ beyond the value $D/J \simeq 10.0$. Accordingly, it is intuitive to roughly separate the observed short-range magnetic behavior of the ac susceptibility into two regions which we preliminarily call the intermediate anisotropy region $6 \leq D/J \leq 10$ and the strong anisotropy region $D/J > 10$, respectively. One question of our concern for these anisotropy regions is that whether there exhibits a critical slowing down scaling law as described in Eq. (4) so that a spin-glass transition temperature $T_g$ to some magnetic glassy phase still does exist. To address this question, we perform our study of the ac susceptibility for another case of intermediate anisotropy strength $D/J = 10$ similar to that of the $D/J = 3.5$ case. Figure 3(b) presents the temperature dependence of $\chi'(T,\omega)$ for the $D/J = 10$ case at various frequencies. Apparently, the dispersionless peak is not observed for all of the frequencies while the high-temperature peak still shifts to low temperature with decreasing frequency similar to what we observed in Fig. 2(a) for the $D/J = 3.5$ case. To our surprise, the plot of $J/T_b(\omega)$ vs. $\log(1/\omega)$, as shown in the inset of Fig. 3(b), indicates a conventional Arrhenius law (i.e., a straigthline dependence), which is quite similar to the behavior of an ensemble of noninteracting nonoparticles, and no such a critical slowing down law as given by Eq. (4) is fulfilled. The absence of the critical slowing down of a magnetic glassy transition, however, contrasts to some experimental results, namely that recently observed by Luo *et al.* [32] for the Dy-based bulk magnetic glass $Dy_{40}Al_{24}Co_{20}Y_{11}Zr_5$, a strong random anisotropy magnet. The reason is likely due to that the frequencies in our simulation are restricted to vary in a so narrow range (one order of magnitude) that only a linear relationship of $J/T_b(\omega)$ vs. $\log(1/\omega)$ is observed. Further result for a wider range of frequency should be provided to have a conclusive reasoning.

In conclusion, our Monte Carlo simulation gives feacture characteristic of magnetic ordering and collective dynamics for the case $D/J = 3.5$. With the hypothesis of cluster formation, the simulated results show a "smeared" SP-to-SRO transition at $T_C \approx 1.5$, followed by an onset of magnetic phase transition at $T_g \approx 1.2$ below which both the correlated spin-glass and magnetic quasi-long-range order coexist. We demonstrated that the features for the $D/J = 3.5$ case are typically characteristic of those of systems possessing weak anisotropy strengths in the range $1.0 \leq D/J \leq 5.0$. When this range is exceeded, the QLRO is predicted to be totally suspended while whether a magnetic glassy transition temperature $T_g$ does exist is still an open question for further investigations. Our study supports a recent theoretical point of view that QLRO is as a rare and novel order that distinguishes the weak random-anisotropy glasses from those of strong random anisotopy [29].

This work was financially supported by the National Science Council (NSC) of Taiwan, R.O.C, under Contract No. NSC 95-2112-M-007-025-MY2. The computing resources of the National Center for High-performance Computing under project "Taiwan Knowledge Innovation National Grid" are acknowledged.

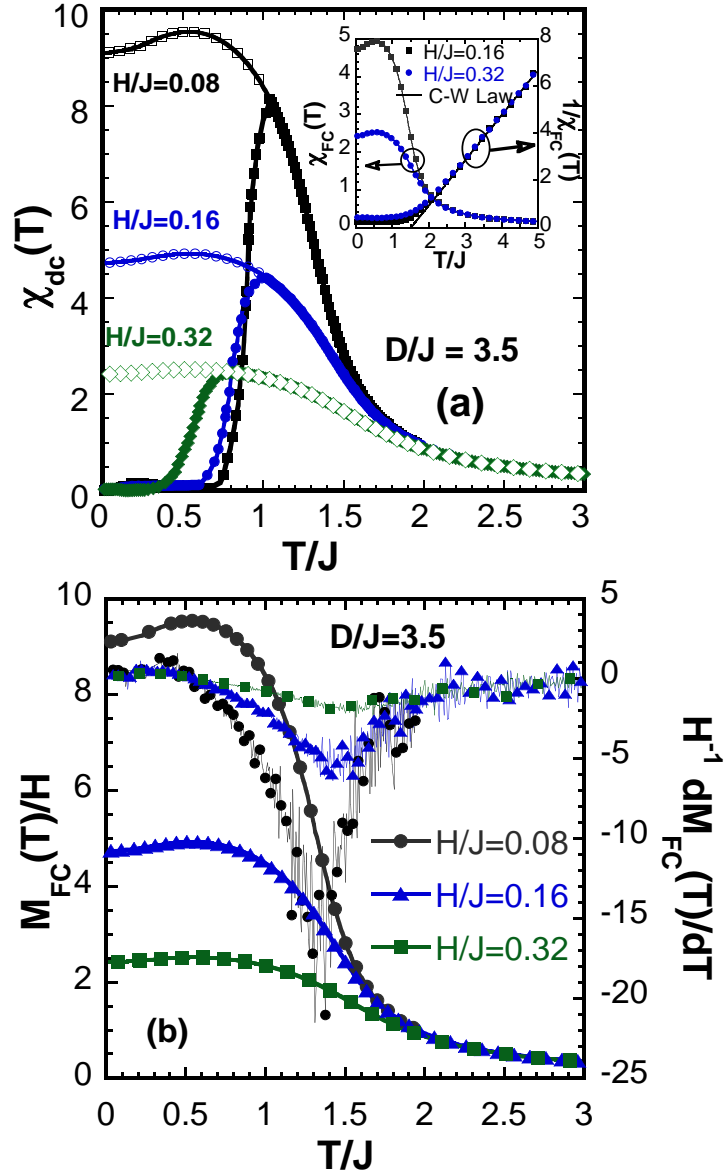

FIG. 1: DC magnetic measurements for both the ZFC and the FC protocols with various applied fields for the $D/J = 3.5$ case. (a) $\chi_{\text{ZFC}}(T)$ and $\chi_{\text{FC}}(T)$; the best linear fit for $T/J > 2.0$ (inset) indicates a Curie-Weiss-like tranistion at $T_{\text{C}}/J \approx 1.5$. (b) $M_{\text{FC}}(T)/H$ and $H^{-1}(\partial M_{\text{FC}}(T)/\partial T)$.

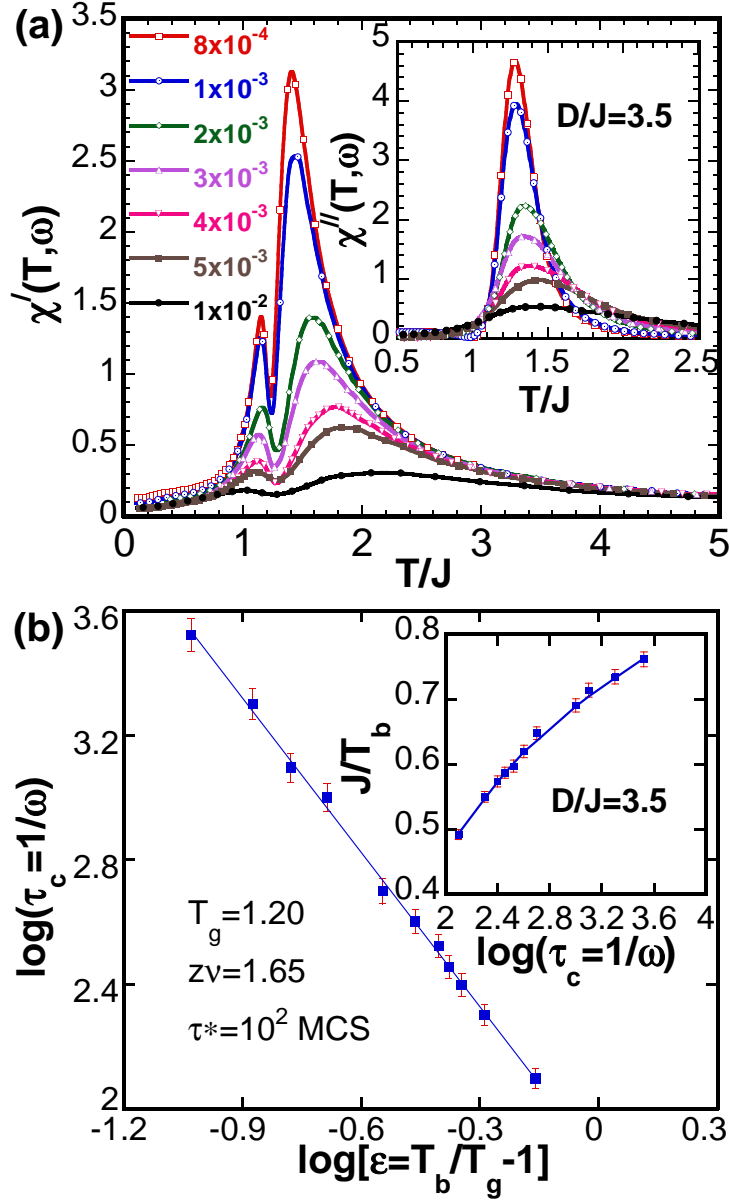

FIG. 2: Simulated results for the temperature dependence of the ac susceptibility at various frequencies for $D/J = 3.5$. (a) $\chi'(T,\omega)$ and $\chi''(T,\omega)$ (inset). (b) The critical slowing-down scaling $\log(\tau_c = 1/\omega)$ vs. $\log(\varepsilon = 1 - T_b/T_g)$ yields $T_g \approx 1.20$ and $z\nu \approx 1.65$ and the Arrhenius plot of $J/T_b$ vs. $\log(\tau_c = 1/\omega)$(inset) indicates a non-Arrhenius behavior.

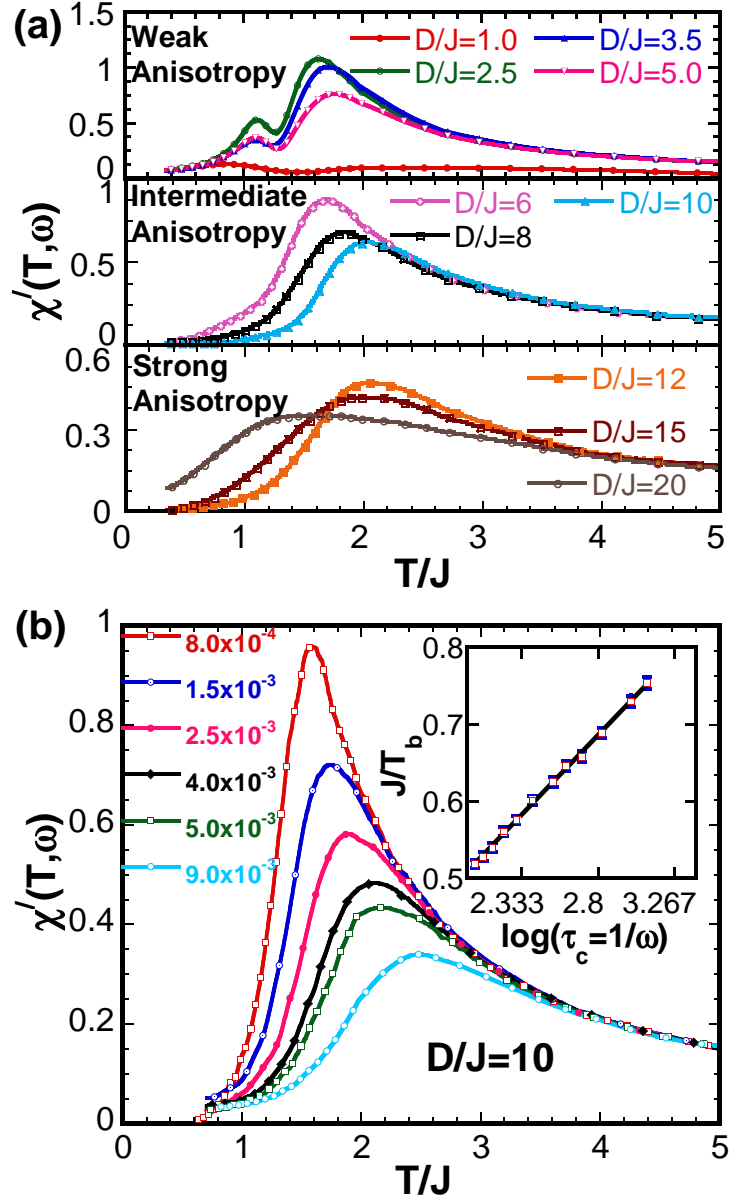

FIG. 3: Clues for boundaries among anisotropy regions. (a) $\chi'(T,\omega)$ at $\omega = 3 \times 10^{-3}$ (MCS$^{-1}$) for various $D/J$ values. (b) $\chi'(T,\omega)$ at various $\omega$ values. The Arrhenius plot in the inset indicates a regular Arrhenius behavior for the $D/J = 10.0$ case.

14